\documentclass[pra,twocolumn,showpacs,superscriptaddress,nofootinbib]{revtex4}
\usepackage{graphicx}

\usepackage{bm}
\usepackage{amssymb}
\usepackage{amsmath}
\usepackage{amstext}
\usepackage{latexsym}
\usepackage[colorlinks,linkcolor=red]{hyperref}

\begin{document}

\title{Long-time behavior of many-particle quantum decay}

\author{A. del Campo}
\affiliation{Institut f{\"u}r Theoretische Physik, Leibniz Universit\"at Hannover, Appelstrasse 2 D-30167,
Hannover, Germany}

\def\G{\Gamma}
\def\L{\Lambda}
\def\la{\lambda}
\def\g{\gamma}
\def\al{\alpha}
\def\s{\sigma}
\def\e{\epsilon}
\def\k{\kappa}
\def\ve{\varepsilon}
\def\l{\left}
\def\r{\right}
\def\te{\mbox{e}}
\def\d{{\rm d}}
\def\t{{\rm t}}
\def\K{{\rm K}}
\def\N{{\rm N}}
\def\H{{\rm H}}
\def\la{\langle}
\def\ra{\rangle}
\def\om{\omega}
\def\Om{\Omega}
\def\vep{\varepsilon}
\def\wh{\widehat}
\def\tr{\rm{Tr}}
\def\da{\dagger}
\def\iz{\left}
\def\zi{\right}
\newcommand{\beq}{\begin{equation}}
\newcommand{\eeq}{\end{equation}}
\newcommand{\beqa}{\begin{eqnarray}}
\newcommand{\eeqa}{\end{eqnarray}}
\newcommand{\intf}{\int_{-\infty}^\infty}
\newcommand{\into}{\int_0^\infty}

\begin{abstract}

While exponential decay is ubiquitous in Nature,
deviations at both short and long times are dictated by quantum mechanics. 
Non-exponential decay is known to arise due to the possibility of reconstructing the initial state from the decaying products.
We discuss the quantum decay dynamics by tunneling of a many-particle system,
characterizing the long-time non-exponential behavior of the non-escape and survival probabilities. 
The effects of contact interactions and quantum statistics are described. 
It is found that whereas for non-interacting bosons the long-time decay follows a power-law with an exponent linear 
in the number of particles $\N$, the exponent becomes quadratic in $\N$ in the fermionic case. 
The same results apply to strongly interacting many-body systems related by the generalized Bose-Fermi duality.
The faster fermionic decay can be traced back to the effective hard-core interactions between particles, 
which are as well the decaying products, and exhibit spatial anti-bunching which hinders the reconstruction 
of the initial unstable state. The results are illustrated with a paradigmatic model of quantum decay from a trap allowing leaks by tunneling, 
whose dynamics is described exactly by means of an expansion in resonant states.
\end{abstract}

\maketitle
\section{Introduction}
The exponential decay law of unstable systems is found across all fields of physics. In the description of nuclear stability it was already derived in the early days of quantum mechanics \cite{time0}.
However, it is precisely quantum mechanics that imposes deviations of this law during both the short and long-time decay \cite{Khalfin58,Winter61,Ersak69,FGR78}.
In particular, short-time deviations arise from the finite mean energy of the initial state and are associated with the quantum Zeno effect, which was experimentally demonstrated in \cite{shortexp} and can be exploited for different applications \cite{shortZeno}.
The long-time deviations are exhibited by any physical system, described by a hamiltonian $\hat{h}$ whose spectrum ${\rm sp}(\hat{h})$ is bounded from below.
They  are generally characterized by a power-law decay  $1/t^{\alpha}$ with $\alpha>0$ \cite{FGR78,longtheory}. Elusive for about half a century, the observation of this behavior has been reported in a recent experiment \cite{longexp}.
Following the insight by Ersak, the breakdown of the exponential law can be generally attributed to the possible reconstruction of the unstable state from the decaying products \cite{Ersak69}. If this reconstruction is inhibited, the decay dynamics is governed by an exponential law at all times. This argument was recently sharpened in \cite{reconst} where the interference between the reconstructed and non-reconstructed state was shown to be responsible for short-time deviations. Further, it was noticed that the long-time deviations are indeed due to the state reconstruction in a classical, probabilistic sense.
In spite of the abundant theoretical work on non-exponential decay, its many-particle counterpart is to a good extent unexplored. Indeed, 
for a many-body unstable system the discussion of the decay dynamics in terms of one-body observables is not entirely satisfactory. 
In the light of the reconstruction argument, we might expect  correlations between different particles arising from quantum statistics 
and their interactions to play a crucial role.
The short-time deviations are under current investigation and will be discussed elsewhere.
In this paper, we shall describe the multi-particle tunneling decay, paying particular attention to the long-time non-exponential behavior and identifying the key signatures of quantum statistics and hard-core contact interactions.
The paper is organized as follows. In section II we introduce the many-particle non-escape and survival probabilities, and discuss their computation in bosonized and fermionized systems.
In section III we find their asymptotic behavior in a paradigmatic model of tunneling dynamics using the resonant-state expansion formalism. The robustness of these results is discussed in section IV, and their explanation in terms of state reconstruction is provided in section V. We close with a brief summary and an outlook.

\section{Non-escape and survival probabilities of a multi-particle system}
Consider a many-body system described by a wavefunction $\Psi(x_{1},\dots,x_{\N};t)$, symmetric (bosonic) or antisymmetric (fermionic) with respect to permutation of particles. For simplicity we consider effectively
one-dimensional systems relevant to ultracold gases under tight transverse confinement where the transverse degrees of freedom are frozen (for this to be the case the chemical potential and thermal energy are to be much smaller than the transverse vibrational excitation quantum) \cite{BF}. Let us single out a region of interest $\Delta$, with an associate projector (the characteristic function), $\chi_{\Delta}(\hat{x})$.
We define the $\N$-particle non-escape probability as
\beqa
\mathcal{P}_{\N}(t)&:=&\la\Psi|\!\prod_{n=1}^{\N}\!\chi_{\Delta}(\hat{x}_n)|\Psi\ra\nonumber\\
&=&\int_{\Delta^{\N}}\prod_{n=1}^{\N}\! \d x_n|\Psi(x_{1},\dots,x_{\N};t)|^2,
\nonumber\\
\eeqa
which is the probability for the $\N$ particles to be found simultaneously in the $\Delta$-region.
Indeed, $\mathcal{P}_{\N}$ can be extracted from the atom number distribution $p(n,t)$ used in full counting statistics \cite{FCS}, and defined  as the expectation value
\beqa
p(n,t)=\la\delta(\hat{n}_{\Delta}-n)\ra,
\eeqa
where
$\hat{n}_{\Delta}$ is the density operator in the subspace $\Delta$ and $n\in \mathbb{N}$.    As a matter of fact
\beqa
\mathcal{P}_{\N}(t)=p(n=\N,t),
\eeqa
where for trapped ultracold gases $p(n,t)$ can be experimentally measured as in \cite{PIAB}.
Hence, $\mathcal{P}_{\N}(t)$ is a truly multi-particle observable which should not be confused with the non-escape probability defined with respect to the density,
\beqa
{\rm P}_{\rm N}(t):=\la\Psi|\chi_{\Delta}(x_1)
|\Psi\ra
=\int_{\Delta}\!\d x \,n(x,t),
\eeqa
where
\beqa
n(x,t)=\int_{-\infty}^{\infty}\prod_{n=2}^{\N} \d x_n|\Psi(x_,x_2\dots,x_{\N};t)|^2
\eeqa
is the density profile of the cloud, which imposes no condition on the location of the $(\N-1)$ particles.
The behavior of ${\rm P}_{\rm N}(t)$, the integrated density profile which is a one-body observable, has been the subject of recent studies dealing with dynamics of ultracold gases \cite{DDGCMR06,Ceder09}.

In the description of single-particle quantum decay problems of an unstable state $|\psi\ra$ (where for $\N=1$ the distinction between $\mathcal{P}_{\rm 1}$ and ${\rm P}_{\rm 1}$ becomes superfluous),  the non-escape probability $\mathcal{P}_{\rm 1}(t)$ often behaves in the same way than the fidelity or survival probability $\mathcal{S}_{\rm 1}(t)=|\la\psi(0)|\psi(t)\ra|^2$. The survival probability is referred to the overlap between
the initial and time-evolved state so that its many-particle version simply reads,
\beqa
\mathcal{S}_{\N}(t):=|\la\Psi(0)|\Psi(t)\ra|^2,
\eeqa
and has recently been used to describe Loschmidt echoes in one-dimensional interacting Bose gases \cite{Buljan11}.

\subsection{Fermionized systems\label{FS}} Let us consider an atom number state of a spin-polarized Fermi gas. Since $s$-wave scattering is suppressed by the Pauli exclusion principle, and $p$-wave scattering is generally weak, the Fermi wavefunction of the ground state is well approximated by a Slater determinant,
$\Psi_{F}(x_{1},\dots,x_{\N};t) =\frac{1}{\sqrt{\N!}}{\rm det}_{n,k=1}^{\N}\phi_{n}(x_{k};t)$,
where $\phi_{n}(x,0)$ is the $n-$th eigenstate of the initial Hamiltonian, whose time evolution for $t>0$
following a quench of the trapping potential at $t=0$ is denoted by $\phi_{n}(x,t)$.
Using the Leibniz formula it reads
$\Psi_{F}(x_{1},\dots,x_{\N};t)
=\frac{1}{\sqrt{\N!}}\sum_{P\in S_{\N}}(-1)^P\prod_{n=1}^{\N}\phi_{P(n)}(x_{n},t),$
 where $P$ labels a permutation of the symmetric group $S_{\N}$ with $\N!$ elements
and $(-1)^P=\pm 1$ is the signature of $P$.
This allows us to rewrite the $\N$-particle non escape probability as
\beqa
\label{PNF}
\mathcal{P}_{\rm N}
^{(F)}
(t)\!&=&\!\frac{1}{\N!}\sum_{P,Q\in S_{\N}}\!\!\!(-1)^{P+Q}\!\prod_{n=1}^{\N}
\la\phi_{P(n)(t)}|\chi_{\Delta}|\phi_{Q(n)}(t)\ra\nonumber\\
&=&
\frac{1}{\N!}\sum_{P,R\in S_{\N}}(-1)^{R}\prod_{n=1}^{\N}\la\phi_{n}(t)|\chi_{\Delta}|\phi_{R(n)}(t)\ra\nonumber\\
&=&
{\rm det}_{n,k=1}^{\N} [\la\phi_{n}(t)|\chi_{\Delta}|\phi_{k}(t)\ra],
\eeqa
where $P,Q,R$ label different permutations and \cite{note1}
\beqa
\la\phi_{n}(t)|\chi_{\Delta}|\phi_{k}(t)\ra=\int_{\Delta}\d x \phi_n^*(x,t)\phi_k(x,t) .
\eeqa

Similarly one can compute the survival probability,
\beqa
\label{SNF}
\mathcal{S}_{\rm N}
^{(F)}
(t)
&=&|\la\Psi_F(0)|\Psi_F(t)\ra|^2\nonumber\\
&=&
\left|{\rm det}_{n,k=1}^{\N} [\la\phi_{n}(0)|\phi_{k}(t)\ra]\right|^2,
\eeqa
where ${\rm det}_{n,k=1}^{\N} [\la\phi_{n}(0)|\phi_{k}(t)\ra]$ is the survival amplitude. 
The same expressions in Eq. (\ref{PNF}) and (\ref{SNF}) describe as well hard-core bosons in one dimension, in the so called Tonks-Girardeau (TG) regime \cite{Girardeau60}.
Indeed, their wavefunctions are related by
 the Bose-Fermi mapping
\beqa
\Psi_{TG}(x_{1},\dots,x_{\N})= \mathcal{A}(\hat{x}_{1},\dots,\hat{x}_{\N})\Psi_{F}(x_{1},\dots,x_{\N}),\nonumber
\eeqa
through the antisymmetric unit function
\beqa
\mathcal{A}=\prod_{1\leq j<k\leq \N}\epsilon(\hat{x}_{k}-\hat{x}_{j}),
\eeqa
with  $\epsilon(x)=1$ $(-1)$ if $x>0$ $(<0)$ and $\epsilon(0)=0$ .
Note that this operator is its own inverse
and consequently
\beqa
\mathcal{S}_{\rm N}^{(F)}(t)\equiv\mathcal{S}_{\rm N}^{(TG)}(t)\quad
{\rm  and} \quad \mathcal{P}_{\rm N}^{(F)}(t)\equiv\mathcal{P}_{\rm N}^{(TG)}(t)
\eeqa
are shared by both dual systems, as well as any other multi-particle {\it fermionized } system with hard-core interactions and intermediate statistics, i.e. the whole family of hard-core anyons $\{\Psi_{HCA}^{\theta}=\mathcal{A}_{-\theta}\Psi_{F}|\theta\}$ for which the same mapping holds up to the replacement of $\mathcal{A}$ by $\mathcal{A}_{-\theta}=\prod_{1\leq j<k\leq N}e^{-i\frac{\theta}{2}\epsilon(\hat{x}_{k}-\hat{x}_{j})}$, where $\theta$ is the statistical parameter \cite{Girardeau06,delcampo08}.

\subsection{Bosonized systems\label{BS}} For a non-interacting Bose-gas, the wavefunction of the ground state is built as a Hartree product, $\Psi_{B}(x_{1},\dots,x_{\N};t) =\prod_{n=1}^{\N}\phi_{1}(x_n,t)$,
where $\phi_{1}(x,t)$ describes the time-evolution of the single-particle ground state of the initial Hamiltonian.
It follows that the $\N$-particle non-escape probability becomes the $\N$-th power of the single-particle non-escape probability $\mathcal{P}_1(t)$,
\beqa
\mathcal{P}_{\rm N}^{(B)}
(t)&=&\la\Psi_B|\prod_{n=1}^{\N}\chi_{\Delta}|\Psi_B\ra
=\mathcal{P}_1(t)^{\N},
\eeqa
where the $\mathcal{P}_{\rm 1}(t)=\int_{\Delta}\d x|\phi_1(x,t)|^2$ is referred to evolution of the ground state $\phi_1(x,t)$. Similarly,
$\mathcal{S}_{\rm N}^{(B)}=\mathcal{S}_1^{\N}$, with $\mathcal{S}_{\rm 1}=|\la\phi_1(0)|\phi_1(t\ra|^2$.
These expressions apply generally to {\it bosonized} systems, those related by the Bose-Fermi duality to non-interacting bosons, like the so-called Fermionic Tonks-Girardeau gas and the continuous family of anyonic extensions \cite{Girardeau06,delcampo08}.
Approaching the strictly non-interacting limit in an ultracold atomic Bose gas is somehow challenging.
The same behavior is observed in a weakly interacting gas in the mean-field regime, where the Hartree approximation holds, leading to
\beqa
\mathcal{P}_{\rm N}^{(BEC)}(t)=\mathcal{P}_1(t)^{\N}
\eeqa 
where $\mathcal{P}_1(t)=\int_{\Delta}\d x|\varphi(x,t)|^2$ (and analogously for $\mathcal{S}_{\N}$), with $\varphi(x,t)$ being the solution of the Gross-Pitaevskii equation. Nonetheless, an expanding Bose gas with finite interactions eventually acquires a Tonks-Girardeau structure \cite{OS02,Dario08}.

\section{Model} 
The study of long-time quantum decay becomes extremely challenging or merely intractable with the usual numerical propagation methods based of space-discrete lattices even with the use of complex absorbing potentials in the boundaries of the propagation box \cite{Minguzzi04}.
As a result, scarce analytic results \cite{Winter61,MG09} or specific methods adapted to study quantum decay \cite{FGR78} are required.
We shall use the Resonant-States Expansion (RSE) formalism developed by Garc\'ia-Calder\'on and coworkers \cite{GMM95,DGCM09}, in combination with asymptotically exact expansions.
We choose the one dimensional analogue of the Winter model, one of the paradigmatic models of quantum decay by tunneling \cite{Winter61}.
It consists of an initial box-like potential located in the interval $[0,a]$ with infinite walls for $t<0$ and whose right wall, located at $x=a$,
 is weakened for $t>0$ to a delta function potential
$\hat{v}(x>0)=\eta'\delta(\hat{x}-a)$ of strength $\eta=2m\eta'/\hbar^2>0$, while the left wall $\hat{v}(x\leq 0)$ remains infinite for all $t$, restricting the dynamics to $x>0$.
Its piecewise definition allows for a nearly analytical treatment which facilitates the study of asymptotics both a long and short times.
Let $\Delta:=[0,a)$ be the region of interest.
We shall study the time evolution of the $n$-th eigenstate of the initial box-like trap,
with the general form
$\phi_{n}(x,t=0)=\sqrt{\frac{2}{a}}\sin \left(k_n x\right)\chi_{\Delta}(x)$,
with $k_n=\frac{n\pi}{a}$, $n\in\mathbb{N}$.
For $t>0$, the time evolution of $\phi_n$ reads
\begin{equation}
\label{superposition}
\phi_n(x,t) = \int_{\Delta} g(x,x';t)\phi_n(x',0) \d x',
\end{equation}
where  $g(x,x';t)$ is the retarded Green's function.
The RSE \cite{GMM95,DGCM09} will allow us to exploit the analytical properties of the corresponding outgoing Green's function $G^+(x,x';k)$, after rewriting $g(x,x';t)$  as
\begin{eqnarray}
g(x,x';t)&=& \sum_{j=1}^{\infty} u_j(x)u_j(x')e^{-i\hbar k_j^2t/2m} \nonumber\\
&+& \frac{i}{\pi} \int_{\Gamma}
 G^+(x,x';k)e^{-i\hbar k^2 t/2m} k \d k.
 \label{equg}
\end{eqnarray}
Here, $\{u_j\}$ are the resonant states with complex eigenvalues $E_j=\frac{\hbar^2k_j^2}{2m}=\varepsilon_j-i\gamma_j/2$, lifetime $\hbar/\gamma_j$ and real energy $\varepsilon_j$, which obey outgoing boundary conditions at $x=a$ and satisfy the Schr\"odinger equation. The sum runs over the set of proper complex poles $k_j$ of the scattering S-matrix lying on the fourth quadrant of the complex $k$-plane with ${\rm Re}(k_j)>|{\rm Im} (k_j)|$.
The outgoing Green function $G^+(x,x';k)$ can generally be written explicitly in terms of the regular function and the Jost function of the scattering problem \cite{Newtonbook}. The integral term  involving it, is responsible for  deviations from the exponential law at ultrashort or very long times and we shall focus on its contribution. The path  $\Gamma$ can be chosen as the straight line on the complex $k$-plane ${\rm Im}(k)=-{\rm Re}(k)$, passing through the origin $k=0$ \cite{GMM95}.
This expansion holds as long as $x,x'\in\Delta$, a limitation that can be overcome following \cite{DN99}, should that be necessary. 
Furthermore, in the RSE of $g(x,x';t)$ the sum over resonant states decays exponentially with time.
Rewriting the countor integral as
%
\beqa
&& \frac{i}{\pi} \int_{\Gamma}
 G^+(x,x';k)e^{-i\hbar k^2 t/2m} k \d k \nonumber\\
&& = \frac{(1+i)}{\sqrt{\pi}}\sum_{s=1}^{\infty}\frac{i^{3s}2^{1-s}}{(s-1)!}
G_{xx'}^{(2s+1)}\left(\frac{m}{\hbar t}\right)^{s+\frac{1}{2}}.\nonumber\\
\eeqa
where $G_{xx'}^{(r)}=\partial_k^rG^+(x,x';k)|_{k=0}$, we are ready to 
discuss the long-time behavior of both $\mathcal{P}_{\N}(t)$ and $\mathcal{S}_{\N}(t)$.
The long-time asymptotics of the single-particle wavefunction
is found to be $ \phi_n(x,t)\sim \frac{2(-1)^{1/4}}{\pi^{3/2}}\frac{(-1)^nx}{n(1+\eta a)^2}\left(\frac{ma}{\hbar t}\right)^{3/2}+\mathcal{O}(t^{-5/2})$. 
To leading order, we find the matrix elements
\beqa
\la\phi_{n}(t)|\chi_{\Delta}|\phi_{k}(t)\ra&\propto&\frac{(-1)^{n+k}}{nk(1+\eta a)^4}\left(\frac{ma^2}{\hbar t}\right)^{3},\\
\la\phi_{n}(0)|\phi_{k}(t)\ra&\propto&\frac{(-1)^{n+k}}{nk(1+\eta a)^2}\left(\frac{ma^2}{\hbar t}\right)^{3/2},
\eeqa
so that for non-interacting bosons
\beqa
\label{longB}
\mathcal{P}_{\N}^{(B)}(t)\propto \mathcal{S}_{\N}^{(B)}(t)\propto \frac{1}{(1+\eta a)^{4\N}}\left(\frac{ma}{\hbar t}\right)^{3\N}.
\eeqa
Hence, the exponent of the power-law is linear in the particle number $\N$.
The late exponential quantum decay preceding the regime in Eq. (\ref{longB}) is governed by the resonance with longest life time $\gamma_1$, leading to a multiparticle decay rate
\beqa
\Gamma_{\rm N}=-d\ln\mathcal{P}_{\rm N}(t)/dt\simeq\N\gamma_1.
\eeqa
Deviations set in during the crossover
\beqa
\exp(-\Gamma_{\N}t)\approx\mathcal{P}_{\rm 1}^{\N}
\eeqa
characterized by a non-monotonic behavior, and hence the onset of the $\mathcal{P}_{\rm N}$ ($\mathcal{S}_{\N}$) asymptotics is the same as in the single-particle dynamics, and can be characterized by the single resonant criterion, which demands a small ratio $R=\varepsilon_{1}/\gamma_1$ for deviations to occur \cite{R}.

In the case of fermions ($\N>1$) the first order of the expansion vanishes due to
symmetry in Eqs. (\ref{PNF}) and (\ref{SNF}) to $\mathcal{O}(t^{-3N})$, and higher order terms are to be taken into account.
The exact expression of $G^+(x,x';k)$ in the 1D analogue of the Winter model can be constructed \cite{GCMV07}.
One can use it to compute the matrix elements entering in the definition of $\mathcal{S}_{\N}$ and $\mathcal{P}_{\N}$ and find the leading contribution for different $\N$, which takes the form \cite{note2}
\beqa
\label{nelongF}
\mathcal{P}_{\rm 1}&\sim&\frac{4 }{3 \pi^3}\frac{1}{(1+\eta a)^{4}} \left(\frac{ma^2}{\hbar  t}\right)^{3}\!,\nonumber\\
\mathcal{P}_{\rm 2}^{(F)}&\sim& \frac{3 }{175  \pi^{10} }\frac{1}{(1+\eta a)^{8}}\left(\frac{ma^2}{\hbar  t}\right)^{10}\!,
\nonumber\\
\mathcal{P}_{\rm 3}^{(F)}&\sim& \frac{1024 }{6015380679 \pi ^{21}}\frac{1}{(1+\eta a)^{12}}\left(\frac{ma^2}{\hbar  t}\right)^{21}\!,
\dots\nonumber\\
\mathcal{P}_{\rm \N}^{(F)}&\propto& \frac{1}{(1+\eta a)^{4\N}}\left(\frac{ma^2}{\hbar  t}\right)^{\N(2\N+1)}\!,
\eeqa
where the scaling in the last equation has been verified for finite values of $\N$ by explicit expansion of $\mathcal{P}_{\rm \N}$.
The survival probability exhibits the same long-time asymptotic law,
the first few terms being
\beqa
\mathcal{S}_{\rm 1}&\sim&\frac{8 }{ \pi^5}\frac{1}{(1+\eta a)^{4}}\left(\frac{ma^2}{\hbar  t}\right)^{3},\nonumber\\
\mathcal{S}_{\rm 2}^{(F)}&\sim& \frac{729 }{16  \pi^{18}}\frac{1}{(1+\eta a)^{8}}\left(\frac{ma^2}{\hbar  t}\right)^{10}, \nonumber\\
\mathcal{S}_{\rm 3}^{(F)}&\sim& \frac{8000000 }{531441 \pi ^{39}}\frac{1}{(1+\eta a)^{12}}\left(\frac{ma^2}{\hbar  t}\right)^{21},\dots
\nonumber\\
\mathcal{S}_{\rm \N}^{(F)}&\propto& \frac{1}{(1+\eta a)^{4\N}}\left(\frac{ma^2}{\hbar  t}\right)^{\N(2\N+1)}.
\eeqa
The upshot is that for spin-polarized fermions, and fermionized systems such as the Tonks-Girardeau gas, 
the exponent of the long-time power-law decay becomes quadratic (instead of linear) in the number of particles $\N$.

\section{Stability of the scaling} The leading contribution of every term in the Leibniz expansion of the determinant formula of the non-escape probability goes as $1/t^{3}$, and as $1/t^{3/2}$ in the case of the survival amplitude. The appearance of any power-law other than that in Eq. (\ref{longB}) results from the cancelation of this leading term by symmetry, and the contribution of higher-order terms.
The question arise as to whether the symmetry of the system is responsible for the multi-particle scaling.
The asymptotic behavior is robust against a finite shift of the potential barrier to a position $d\neq a$, $d<\infty$, up to the numerical factors and the role played by the initial width of the cloud $a$. For bosons Eq. (\ref{longB}) holds with
\beqa
\mathcal{P}_{\rm \N}^{(B)}\propto\mathcal{S}_{\rm \N}^{(B)}\propto\frac{1}{(1+\eta d)^{4\N}}\left(\frac{mad}{\hbar  t}\right)^{3\N},
\eeqa
while in the fermionic case
\beqa
\mathcal{P}_{\rm \N}^{(F)}\propto\mathcal{S}_{\rm \N}^{(F)}\propto \frac{1}{(1+\eta d)^{4\N}}\left(\frac{mad}{\hbar  t}\right)^{\N(2\N+1)}.
\eeqa

The limit of a vanishing barrier is non-singular and can be solved analytically, see Appendix A. The case $\N=2$ was recently considered 
in \cite{TS11}, see also \cite{MG11}.

\section{Discussion}
One might be tempted to conclude that the faster long-time decay of fermionized systems in comparison with bosonized gases results from the higher mean energy of the former.
To appreciate that this is not the case, consider an excited state of $\N$ bosons degenerate with the ground state of the non-interacting spin-polarized Fermi gas, with one single atom in each single-particle eigenstate,
$
 \Psi^{(EB)}(x_{1},\dots,x_{\N};t)
=
\frac{1}{\sqrt{\N!}}{\rm per}_{n,k=1}^{\N}\phi_{n}(x_{k};t)
=\frac{1}{\sqrt{\N!}}\sum_{P\in S_{\N}}\prod_{n=1}^{\N}\phi_{P(n)}(x_{n},t)$,
$EB$ standing for excited Bose gas.
The corresponding non-escape and survival probabilities
can be found to be
\beqa
\mathcal{P}_{\rm N}^{(EB)}(t)
&=&{\rm per}_{n,k=1}^{\N} [\la\phi_{n}(t)|\chi_{\Delta}|\phi_{k}(t)\ra],\\
\mathcal{S}_{\rm N}^{(EB)}(t)
&=&\left|{\rm per}_{n,k=1}^{\N} [\la\phi_{n}(0)|\phi_{k}(t)\ra]\right|^2.
\eeqa
The density profile of fermions and EB becomes indistinguishable
\beqa
n^{(F)}(x,t)=n^{(EB)}(x)=\sum_{k=1}^{\N}|\phi_{k}(x,t)|^2,
\eeqa
whence it follows that these systems share the same one-body non-escape probability derived from the integrated density profile
\beqa{\rm P}_{\N}^{(F)}\equiv {\rm P}_{\N}^{(EB)}\propto 1/t^3.\eeqa
Nonetheless,
\beqa
\label{longEB}
\mathcal{P}_{\N}^{(EB)}\propto \mathcal{S}_{\N}^{(EB)}\propto1/t^{3\N},
\eeqa
at variance with
\beqa
\mathcal{P}_{\N}^{(F)}\propto \mathcal{S}_{\N}^{(F)}\propto1/t^{\N(2\N+1)}.
\eeqa
Indeed, the same scaling in Eqs. (\ref{longB}) and (\ref{longEB}) holds for thermal states in the canonical ensemble, showing the independence on the mean energy of the initial state.
Moreover, these long-time power-laws are not specific of the 1D Winter model, but actually holds for all potentials in half-space, where the RSE formalism can be applied \cite{GMM95,DGCM09}.

We further note that from the invariance of $\mathcal{P}_{\rm \N}$ and $\mathcal{S}_{\rm \N}$ under Bose-Fermi mappings, it is actually inappropriate to attribute the different scaling to the symmetrization imposed by quantum statistics.
The different asymptotic behavior is indeed a consequence of the hard-core interactions, which lead to fermionization in the sense of  the Bose-Fermi duality \cite{Girardeau06}, not to be confused with its dynamical counterpart \cite{RM05,delcampo08}.
For polarized fermions, the effective hard-core constraint follows from the Pauli exclusion principle.
Either way, these systems exhibits spatial antibunching which hinders the reconstruction of the initial state in comparison with the case of bosonized systems, those dual to non-interacting bosons, and free-of hard-core interactions. An exponential decay law holds for all times in the absence of state reconstruction \cite{Ersak69}, while a slowly decaying asymptotic power-law is favored in those bosonized systems which tend to spatial bunching, hence facilitating the reconstruction of the initial state localized in the trap. Hard-core interactions lead to a faster power-law decay as a result of spatial anti-bunching.

In view of the interaction effects on state reconstruction just discussed, and the results by Taniguchi and Sawada (restricted to free decay in half space of $\N=2$ fermions with both attractive and repulsive Coulomb interactions) \cite{TS11}, we notice that the stages of quantum decay can be further modified whenever the inter-particle potential is of finite range. In the attractive case, the asymptotic power-law can be slowed down with respect to that of bosonized systems, while in the repulsive case long-time deviations can be  suppressed, extending the regime of validity of the exponential decay law.

\section{Conclusions and outlook}
Summing up, we have analyzed the many-particle quantum decay from a leaking trap by tunneling, characterizing the fundamental deviations from the exponential decay law.
The long-time behavior, governed by a power-law in time, exhibits a signature of quantum statistics and short-range contact interactions which drastically modifies the exponent of the power-law as a function of the particle number.
Hard-core interactions lead to spatial antibunching which hinders the initial state reconstruction, and ultimately induces a faster power-law decay than in those systems free of them. We have also shown that this behavior is not related to the higher mean energy of the initial state, whose effect is restricted to the short-time dynamics.
As an outlook, we point out that the use of the exact Bose-Fermi mapping for one-dimensional Bose gases with finite-interactions \cite{Dario08,Buljan08}, might lead to novel asymptotic regimes of tunneling decay, not captured by mean-field theories.

{\it Note-} After the submission of this work, Ref. \cite{copy} has discussed the quantum decay in the same potential model in the two-particle case.

It is a pleasure to acknowledge discussions with J. G. Muga, M. Pons, A. Ruschhaupt and D. Sokolovski, and
to thank the Centre of Excellence for Quantum Engineering and Space-Time Research (QUEST).
\appendix


%
%
%

\section{Free decay in the half-axis}

For a vanishing barrier $\eta\rightarrow0$, the dynamics is free in the positive semiaxis and
a fully analytical solution is available,
writing the propagator by the method of images in terms of the free one \cite{DGCM09}
\beqa
g_{0}(x, x';t)=\sqrt{\frac{m}{2\pi i\hbar t}}
\exp\bigg[\frac{im(x-x')^{2}}{2t\hbar}\bigg]
\eeqa
as
\beqa
\label{prop}
g(x, x';t)=g_{0}(x, x';t)-g_{0}(-x, x';t).
\eeqa
The free expansion in the whole space, under $g_{0}(x, x';t)$ was discussed in \cite{DM06,delcampo08,DGCM09}.
Using it in the superposition principle Eq. (\ref{superposition}) and  Eq. (\ref{prop}), introducing
$\tau=\hbar t /m$, one finds
\beqa
\label{semifree}
\phi_{n}(x,t)&=&\frac{i}{\sqrt{2a}}\!\sum_{\alpha,\beta=\pm1}\alpha\beta
\big[M(\beta x-a,\alpha k_{n},\tau)\nonumber\\
& & +M(\beta x,\alpha k_{n},\tau)\big].
\eeqa
We have used the definition of the Moshinsky function
\beqa
M(x,k,\tau)=\frac{e^{i\frac{x^{2}}{2 \tau}}}{2}w(-z),
\eeqa
where
\beqa
z=\frac{1+i}{2}\sqrt{\tau}\left(k-\frac{x}{\tau}\right),
\eeqa
and  $w$ is the Faddeyeva function
$w(z)= e^{-z^{2}}{\rm{erfc}}(-i z)$ \cite{DGCM09}.

Either using the asymptotic form of the Moshinsky function \cite{DGCM09} or the expansion
\beqa
g(x,x';t)=
-\frac{1+i}{\sqrt{\pi}} \left(\frac{m}{\hbar t}\right)^{3/2} x x'
+\mathcal{O}(t^{-5/2}),
\eeqa
one can recover for bosons the result in Eq. (\ref{longB}).
For fermionized systems one needs to keep higher orders in the expansion, and take into account the slowest power law, which leads to the scaling in Eq. (\ref{nelongF}).


\begin{thebibliography}{99}
\bibitem{note2} The power-law  governing the
long-time asymptototics of $\mathcal{P}_1(t)\propto t^{-\alpha}$ does change with finite range potentials. In particular, the values of $\alpha$ for $v(x)\sim q_0/x^p$ as $x\rightarrow\infty$ with $p\in[2,3)$ are known \cite{longtheory}. $\mathcal{P}_N(t)\propto t^{-\alpha \N}$ holds only as long as $\alpha$ does not depend on the quantum number $n$ of the initial state. Short-range potentials in 3D including a centrifugal term $p=2$, lead to $\alpha=\alpha_n$. Generally, for angular momenta $l$ other than zero, the outgoing  Green function acquires a $l$-dependence $G^+=G_l^+$.
\end{thebibliography}
\end{document}